\documentclass[twocolumn,showpacs,preprintnumbers,superscriptaddress,amsmath,amssymb]{revtex4}
  
\usepackage{graphicx}  % needed for figures
\usepackage{dcolumn}   % needed for some tables
\usepackage{bm}        % for math
\usepackage{amssymb}   % for math

\def\missET {{\not\!\! E_T}}

\begin{document}

\title{Cross Section Constrained Top Quark Mass \\ Measurement from Dilepton Events at the Tevatron}% Force line breaks with \\
 
\affiliation{Institute of Physics, Academia Sinica, Taipei, Taiwan 11529, Republic of China} 
\affiliation{Argonne National Laboratory, Argonne, Illinois 60439} 
\affiliation{Institut de Fisica d'Altes Energies, Universitat Autonoma de Barcelona, E-08193, Bellaterra (Barcelona), Spain} 
\affiliation{Baylor University, Waco, Texas  76798} 
\affiliation{Istituto Nazionale di Fisica Nucleare, University of Bologna, I-40127 Bologna, Italy} 
\affiliation{Brandeis University, Waltham, Massachusetts 02254} 
\affiliation{University of California, Davis, Davis, California  95616} 
\affiliation{University of California, Los Angeles, Los Angeles, California  90024} 
\affiliation{University of California, San Diego, La Jolla, California  92093} 
\affiliation{University of California, Santa Barbara, Santa Barbara, California 93106} 
\affiliation{Instituto de Fisica de Cantabria, CSIC-University of Cantabria, 39005 Santander, Spain} 
\affiliation{Carnegie Mellon University, Pittsburgh, PA  15213} 
\affiliation{Enrico Fermi Institute, University of Chicago, Chicago, Illinois 60637} 
\affiliation{Comenius University, 842 48 Bratislava, Slovakia; Institute of Experimental Physics, 040 01 Kosice, Slovakia} 
\affiliation{Joint Institute for Nuclear Research, RU-141980 Dubna, Russia} 
\affiliation{Duke University, Durham, North Carolina  27708} 
\affiliation{Fermi National Accelerator Laboratory, Batavia, Illinois 60510} 
\affiliation{University of Florida, Gainesville, Florida  32611} 
\affiliation{Laboratori Nazionali di Frascati, Istituto Nazionale di Fisica Nucleare, I-00044 Frascati, Italy} 
\affiliation{University of Geneva, CH-1211 Geneva 4, Switzerland} 
\affiliation{Glasgow University, Glasgow G12 8QQ, United Kingdom} 
\affiliation{Harvard University, Cambridge, Massachusetts 02138} 
\affiliation{Division of High Energy Physics, Department of Physics, University of Helsinki and Helsinki Institute of Physics, FIN-00014, Helsinki, Finland} 
\affiliation{University of Illinois, Urbana, Illinois 61801} 
\affiliation{The Johns Hopkins University, Baltimore, Maryland 21218} 
\affiliation{Institut f\"{u}r Experimentelle Kernphysik, Universit\"{a}t Karlsruhe, 76128 Karlsruhe, Germany} 
\affiliation{Center for High Energy Physics: Kyungpook National University, Daegu 702-701, Korea; Seoul National University, Seoul 151-742, Korea; Sungkyunkwan University, Suwon 440-746, Korea; Korea Institute of Science and Technology Information, Daejeon, 305-806, Korea; Chonnam National University, Gwangju, 500-757, Korea} 
\affiliation{Ernest Orlando Lawrence Berkeley National Laboratory, Berkeley, California 94720} 
\affiliation{University of Liverpool, Liverpool L69 7ZE, United Kingdom} 
\affiliation{University College London, London WC1E 6BT, United Kingdom} 
\affiliation{Centro de Investigaciones Energeticas Medioambientales y Tecnologicas, E-28040 Madrid, Spain} 
\affiliation{Massachusetts Institute of Technology, Cambridge, Massachusetts  02139} 
\affiliation{Institute of Particle Physics: McGill University, Montr\'{e}al, Canada H3A~2T8; and University of Toronto, Toronto, Canada M5S~1A7} 
\affiliation{University of Michigan, Ann Arbor, Michigan 48109} 
\affiliation{Michigan State University, East Lansing, Michigan  48824} 
\affiliation{University of New Mexico, Albuquerque, New Mexico 87131} 
\affiliation{Northwestern University, Evanston, Illinois  60208} 
\affiliation{The Ohio State University, Columbus, Ohio  43210} 
\affiliation{Okayama University, Okayama 700-8530, Japan} 
\affiliation{Osaka City University, Osaka 588, Japan} 
\affiliation{University of Oxford, Oxford OX1 3RH, United Kingdom} 
\affiliation{University of Padova, Istituto Nazionale di Fisica Nucleare, Sezione di Padova-Trento, I-35131 Padova, Italy} 
\affiliation{LPNHE, Universite Pierre et Marie Curie/IN2P3-CNRS, UMR7585, Paris, F-75252 France} 
\affiliation{University of Pennsylvania, Philadelphia, Pennsylvania 19104} 
\affiliation{Istituto Nazionale di Fisica Nucleare Pisa, Universities of Pisa, Siena and Scuola Normale Superiore, I-56127 Pisa, Italy} 
\affiliation{University of Pittsburgh, Pittsburgh, Pennsylvania 15260} 
\affiliation{Purdue University, West Lafayette, Indiana 47907} 
\affiliation{University of Rochester, Rochester, New York 14627} 
\affiliation{The Rockefeller University, New York, New York 10021} 
\affiliation{Istituto Nazionale di Fisica Nucleare, Sezione di Roma 1, University of Rome ``La Sapienza," I-00185 Roma, Italy} 
\affiliation{Rutgers University, Piscataway, New Jersey 08855} 
\affiliation{Texas A\&M University, College Station, Texas 77843} 
\affiliation{Istituto Nazionale di Fisica Nucleare, University of Trieste/\ Udine, Italy} 
\affiliation{University of Tsukuba, Tsukuba, Ibaraki 305, Japan} 
\affiliation{Tufts University, Medford, Massachusetts 02155} 
\affiliation{Waseda University, Tokyo 169, Japan} 
\affiliation{Wayne State University, Detroit, Michigan  48201} 
\affiliation{University of Wisconsin, Madison, Wisconsin 53706} 
\affiliation{Yale University, New Haven, Connecticut 06520} 
\author{T.~Aaltonen}
\affiliation{Division of High Energy Physics, Department of Physics, University of Helsinki and Helsinki Institute of Physics, FIN-00014, Helsinki, Finland}
\author{J.~Adelman}
\affiliation{Enrico Fermi Institute, University of Chicago, Chicago, Illinois 60637}
\author{T.~Akimoto}
\affiliation{University of Tsukuba, Tsukuba, Ibaraki 305, Japan}
\author{M.G.~Albrow}
\affiliation{Fermi National Accelerator Laboratory, Batavia, Illinois 60510}
\author{B.~\'{A}lvarez~Gonz\'{a}lez}
\affiliation{Instituto de Fisica de Cantabria, CSIC-University of Cantabria, 39005 Santander, Spain}
\author{S.~Amerio}
\affiliation{University of Padova, Istituto Nazionale di Fisica Nucleare, Sezione di Padova-Trento, I-35131 Padova, Italy}
\author{D.~Amidei}
\affiliation{University of Michigan, Ann Arbor, Michigan 48109}
\author{A.~Anastassov}
\affiliation{Rutgers University, Piscataway, New Jersey 08855}
\author{A.~Annovi}
\affiliation{Laboratori Nazionali di Frascati, Istituto Nazionale di Fisica Nucleare, I-00044 Frascati, Italy}
\author{J.~Antos}
\affiliation{Comenius University, 842 48 Bratislava, Slovakia; Institute of Experimental Physics, 040 01 Kosice, Slovakia}
% ...
\collaboration{CDF Collaboration\footnote{With visitors from $^a$University of Athens, 15784 Athens, Greece, 
$^b$Chinese Academy of Sciences, Beijing 100864, China, 
$^c$University of Bristol, Bristol BS8 1TL, United Kingdom, 
$^d$University Libre de Bruxelles, B-1050 Brussels, Belgium, 
$^e$University of California Irvine, Irvine, CA  92697, 
$^f$University of California Santa Cruz, Santa Cruz, CA  95064, 
$^g$Cornell University, Ithaca, NY  14853, 
$^h$University of Cyprus, Nicosia CY-1678, Cyprus, 
$^i$University College Dublin, Dublin 4, Ireland, 
$^j$University of Edinburgh, Edinburgh EH9 3JZ, United Kingdom, 
$^k$University of Heidelberg, D-69120 Heidelberg, Germany, 
$^l$Universidad Iberoamericana, Mexico D.F., Mexico, 
$^m$University of Manchester, Manchester M13 9PL, England, 
$^n$Nagasaki Institute of Applied Science, Nagasaki, Japan, 
$^o$University de Oviedo, E-33007 Oviedo, Spain, 
$^p$Queen Mary, University of London, London, E1 4NS, England, 
$^q$Texas Tech University, Lubbock, TX  79409, 
$^r$IFIC(CSIC-Universitat de Valencia), 46071 Valencia, Spain, 
}}
\noaffiliation

\date{\today}% It is always \today, today,
             %  but any date may be explicitly specified

\begin{abstract}

We report the first top quark mass measurement that uses a cross section constraint to improve the mass determination. This measurement is made with a dilepton $t\bar{t}$ event sample collected with the CDF II detector. From a data sample corresponding to an integrated luminosity of 1.2~fb$^{-1}$, we measure a top quark mass of $\rm{170.7^{+4.2}_{-3.9}(stat)\pm2.6(syst)}$ $\rm{\pm2.4(theory)\ GeV/{\it{c}}^{2}}$. The measurement without the cross section constraint results in a top quark mass of $\rm{169.7^{+5.2}_{-4.9}(stat)\pm3.1(syst)\ GeV/{\it{c}}^{2}}$.

\end{abstract}

\maketitle

The top quark is the heaviest known elementary particle. Its mass $M_{t}$ is a fundamental parameter in the standard model (SM). Together with the $W$-boson mass, $M_{t}$ places constraints on the SM Higgs boson mass \cite{higgs}. At hadron colliders, the top quarks are mainly pair produced via the strong interaction. Each top quark decays into a $W$-boson and a $b$-quark, and in the dilepton channel both $W$-bosons decay to a charged lepton and a neutrino. The $t\bar{t}$ dilepton events have a small branching ratio, but they have a higher purity than single-lepton or all-hadronic final states. Because the two neutrinos  in the final state are not detected, the dilepton channel top mass fit is under-constrained. However, measuring the mass in this channel is important because it provides an independent measurement of $M_{t}$ that can be compared to measurements in other decay channels, allowing a consistency check of the $t\bar{t}$ hypothesis in the dilepton channel. Previous measurements of $M_t$ in the dilepton channel are described in~\cite{PRD,madcow,dzero_dil_mtop}.

According to the SM, the theoretical $t\bar{t}$ cross section $\sigma_{t\bar{t}}$ has an exponential dependence on the top mass \cite{xsec,fform}. Therefore, the top mass value can be extracted from the observed event yield alone. By combining the theoretical $\sigma_{t\bar{t}}(M_t)$ dependence with the top mass determination from the event kinematics, we can use the cross section information to improve the mass measurement, as reported in this Letter.

In this novel measurement, the constraint provided by the mass dependent theoretical $t\bar{t}$ cross section is combined with a ``template method'' in which a top quark mass $m_t^{rec}$ is reconstructed for each event and in which the distribution of $m_t^{rec}$ is compared with template distributions derived from simulation. In principle, any measurement of $M_t$ and $\sigma_{t\bar{t}}$ could be combined with the theoretical $\sigma_{t\bar{t}}(M_t)$ to obtain an improved top mass determination. However, in our analysis we include the cross section constraint while properly taking into account the top mass dependence of the acceptance and all the correlated systematic uncertainties.

%Unlike a simple $\chi^2$ fit from measured $M_t$, $\sigma_{t\bar{t}}$, and theoretical $\sigma_{t\bar{t}}(M_t)$, this method takes into account that the acceptance of $t\bar{t}$ events depends on the top mass, and properly accounts for correlations between uncertainties.

The template method adopted here is an enhanced version of the ``full kinematic analysis'' described in  \cite{PRD}. The enhanced version treats $b$-tagged and non-tagged events separately. This separation improves the expected statistical uncertainty by 20$\%$; this represents a significant improvement over the previous analysis, which handled $b$-tagged and non-tagged events as a single sample. Introducing the cross section constraint improves the expected statistical uncertainty further by 20$\%$. In this paper, the measurement without the cross section constraint will be referred to as the ``traditional'' measurement. Both measurement techniques were fixed before the data were revealed.

This measurement uses data collected by the CDF II detector corresponding to an integrated luminosity of 1.2~fb$^{-1}$. The CDF II detector \cite{CDF} is a multi-purpose particle detector at the $p\bar{p}$ Tevatron Collider. Charged particle trajectories are measured with a silicon micro-strip detector and a drift chamber, which are immersed in a 1.4~T uniform magnetic field parallel to the beam directions. Electron, photon and hadron energies are measured with electromagnetic and hadronic calorimeters. Muons are detected with drift chambers and scintillation counter hodoscopes located outside the calorimeters.  CDF employs cylindrical coordinates where $\theta$ and $\phi$ are the polar and azimuthal angles with respect to the proton beam. Transverse energy and momentum are defined as $E_T=E\sin\theta$ and $p_T=p\sin\theta$, where $E$ is the energy and $p$ is the momentum.

The data for this analysis were collected using an inclusive lepton trigger that required an electron or a muon with $p_T>$18~GeV/{\it{c}}~\cite{ele_pt}. After the offline reconstruction, the dilepton events were selected using the DIL selection~\cite{DIL}, which requires two oppositely charged leptons with $p_T>$20~GeV/{\it{c}}~\cite{ele_pt}, missing transverse energy \cite{PRD}  due to the undetected neutrinos ($\missET>$~25~GeV), and at least two jets with $E_T>$~15~GeV. The main background sources are diboson and Drell-Yan production and $W$+jets events where a jet is mis-reconstructed as a lepton. Additional cuts were applied to reduce the background~\cite{DIL}.  The expected and observed numbers of events are summarized in Table~\ref{tab:event_sel_exp}. After the event selection, the sample was divided into two subsamples with significantly different signal-to-background ratios. The $b$-tagged sample includes 32 events in which at least one of the jets is identified as a $b$-quark candidate through the presence of a displaced vertex \cite{btagging}. This subsample has an expected signal-to-background ratio of ~11:1. The non-tagged sample comprises 45 events in which none of the jets is identified as a $b$-quark candidate. In this subsample the expected signal-to-background ratio is ~1:1.

\begin{table}
%\centering
\caption{Expected and observed number of events passing event selection criteria. Statistical and systematic uncertainties have been combined.}
\begin{center}
\begin{tabular}{l c }
\hline
\hline
& Expected background\\
diboson & 5.8 $\pm$ 0.9\\
$Z/\gamma^{*}\rightarrow ll, l=e,\mu,\tau$ & 10.9 $\pm$ 2.3\\
misidentified leptons & 8.8 $\pm$ 3.9\\
Total & 25.6 $\pm$ 5.5\\
\hline
& Expected signal \\
$t\bar{t}\ (M_{t}=170\ {\rm GeV}/c^2)$  & 62.1 $\pm$ 4.3 \\
\hline
Total expected & 87.7 $\pm$ 8.9\\
Data & 77\\
\hline
\hline
\end{tabular}
\end{center}
\label{tab:event_sel_exp}
\end{table}

Because the two neutrinos are not detected, the reconstruction of the top quark mass from dilepton events is under-constrained. Top mass reconstruction can be accomplished by considering a kinematic variable which is not observable on an
event-by-event basis, but the distribution of which is predictable and
independent of the top mass value. In this analysis the distribution of $p_z^{t\bar{t}}$, the longitudinal momentum of the $t\bar{t}$ system, was adopted as the variable. Monte Carlo simulations, generated with {\sc pythia}~\cite{pythia} and the CDF II detector simulation~\cite{detsim}, indicate that the distribution of $p_z^{t\bar{t}}$ is nearly independent of the top mass, and is described by a Gaussian distribution with a mean of zero and width of 195~GeV/{\it{c}}. The validity of the Monte Carlo simulation was tested with data from the lepton+jets decay channel where $p_z^{t\bar{t}}$ can be explicitly reconstructed. 

For each event, a top mass $m_t^{rec}$ is reconstructed from the event kinematics as follows. The jet energies are corrected for defects in the calorimeter response, computed from Monte Carlo simulations, to correspond to the energies of the primary $b$-quarks. After these jet energy corrections, the two components of transverse missing momentum are taken as the sum of the neutrino transverse momentum components. Along with assumptions on the masses of the final state particles and  additional constraints on $M_{W^{\pm}} = 80.4\ {\rm{GeV}/c^2}$, $M_{t} = M_{\bar{t}}$, and $p_z^t+p_z^{\bar{t}} = p^{t\bar{t}}_z$, a top mass can be calculated~\cite{PRD}.

A wide range of possible $p_z^{t\bar{t}}$ values is incorporated by calculating the top mass 10000 times. For each iteration, $p_z^{t\bar{t}}$ is randomly drawn from its expected distribution. Similarly, the jet energies and $\missET$ are smeared according to their resolutions. For each iteration, if a solution is not found using the fixed values of $M_W$ and $M_t$, solutions within $M_{W^{\pm}}=80.4\pm3.0$~GeV/$c^2$ and $M_t=M_{\bar{t}}\pm2.0$~GeV/$c^2$ are accepted.

For a given event, we obtain two distributions of possible top quark masses, each corresponding to a different lepton-jet pairing. The pairing which has the higher fraction of solutions is selected. This choice is correct for 70$\%$ of simulated $t\bar{t}$ events. If the number of entries in this distribution is less than 100, the event is rejected. According to Monte Carlo studies, 91$\%$ of signal and 78$\%$ of background events pass this mass reconstruction requirement. The most probable value of a spline fit to the distribution selected is taken as a per-event top mass $m_t^{rec}$. 

Templates of reconstructed top mass distributions were created from various $t\bar{t}$ and background samples. Signal templates were generated from $t\bar{t}$ Monte Carlo samples with generated top masses ranging from 150 to 200~GeV/$c^2$, separately for $b$-tagged and non-tagged signal events. Diboson and $Z\rightarrow ll$ templates were generated from Monte Carlo simulation. A template for misidentified leptons was created using data. The background templates were combined according to the expected contribution of each background source. It was observed from simulation that using the same common background template for $b$-tagged and non-tagged samples provides as good a performance as using separate templates. The common background template was therefore used for both subsamples. 

In the traditional measurement, the top mass is extracted by comparing the reconstructed mass distributions from data to the signal and background template parametrizations ($f_s$ and $f_b$, respectively) using an unbinned likelihood fit. The likelihood includes free parameters for the number of signal events $n_{\rm s}$ and background events $n_{\rm b}$ in each subsample, and for the top mass $M_t$. The total likelihood takes the form

\begin{equation}
{\cal L}  \equiv {\cal L}_{\rm b-tagged}(M_{t},n_{\rm s}^{\rm b},n_{\rm b}^{\rm b}) \times {\cal L}_{\rm non-tagged}(M_{t},n_{\rm s}^{\rm non},n_{\rm b}^{\rm non}) \\
\end{equation}

\noindent where each of the subsample likelihoods is:

\begin{equation}
\begin{array}{rll}
%\begin{eqnarray*}
{\cal L}_{{\rm non/}b{\rm -tagged}} & \equiv & {\cal L}_{\rm shape} \times {\cal L}_{\rm nev} \times {\cal L}_{\rm bg} \\
\vspace{+2mm}
{\cal L}_{\rm shape} & \equiv &  \prod_{i=1}^N \frac{n_{\rm s} \times
 f_{\rm s}(m_{t_i}^{rec},M_{t}) + n_{\rm b} \times
f_{\rm b}(m_{t_i}^{rec})}{n_{\rm s} + n_{\rm b}}\\
\vspace{+2mm}
{\cal L}_{\rm nev} & \equiv & \frac{e^{-(n_{\rm s}+n_{\rm b})}(n_{\rm s}+n_{\rm b})^{N}}{N!}\\
\vspace{+2mm}
-\ln {\cal L}_{\rm bg} & \equiv & \frac{(n_{\rm b} - n_{\rm b}^{\rm exp})^{2}}{2\sigma_{n_{\rm b}}^{2}}.
\end{array}
\label{eq_likelihood}
\end{equation}
%\end{eqnarray*}

\noindent In the formula, $N$ is the number of data events in the subsample, $n_{\rm b}^{\rm exp}$ and $\sigma_{n_{\rm b}}$ are the expected number of background events and its uncertainty, and $m_{t_i}^{rec}$ is the value of $m_t^{rec}$ for the i'th event. The top quark mass hypothesis which minimizes -ln(${\cal L}$) is taken.

To test the method, we performed Monte Carlo experiments of signal and background events. The numbers of signal and background events in each experiment were varied according to Poisson distributions using the expected mean numbers of events. According to the Monte Carlo experiments, the method is unbiased and returns appropriate uncertainties.

\begin{figure}[t]
\begin{center}
\includegraphics[width=0.4\textwidth]{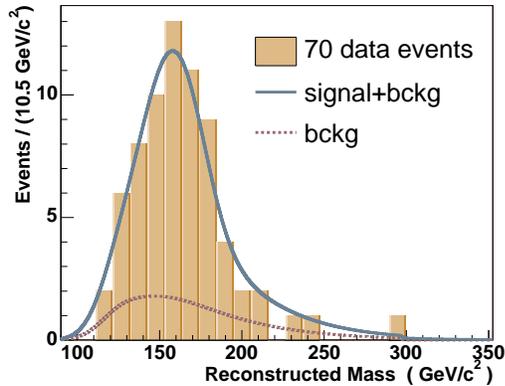}
\caption{Reconstructed top mass distribution from data together with the signal and background parametrizations.}\label{fig:datadistribution}
\end{center}
\end{figure}

In 1.2~fb$^{-1}$ of data, 31 $b$-tagged and 39 non-tagged events pass the event selection criteria and have solutions for $m_t^{rec}$. The mass measurement from the $b$-tagged sample gives $M_{t}=171.9^{+7.2}_{-6.4}$(stat.)~GeV/$c^2$, and the measurement from the non-tagged sample gives $M_{t}=166.0^{+8.4}_{-7.9}$(stat.)~GeV/$c^2$. Applying the traditional method to the two subsamples with a joint likelihood, we measure $M_{t}=169.7^{+5.2}_{-4.9}$(stat.)~GeV/$c^2$. In 34$\%$ of Monte Carlo experiments, the statistical uncertainty was smaller than the measured statistical uncertainty from data. The reconstructed top mass distribution from data is shown in Fig.~\ref{fig:datadistribution}.

%\begin{figure}[htbp]

The top mass measurement can be further improved by taking into account the top mass dependence of the $t\bar{t}$ production cross section. The expected number of signal events can be expressed as

\begin{equation}
n_{s}(M_{t})=\sigma_{t\bar{t}}(M_{t})\cdot a(M_{t})\cdot{L}\cdot p_{mass}^{rec},
\label{eq_ns}
\end{equation}

\noindent where $\sigma_{t\bar{t}}(M_{t})$ is the theoretical $t\bar{t}$ cross section, $a(M_{t})$ is the acceptance of $t\bar{t}$ events, ${L}$ is the integrated luminosity, and $p_{mass}^{rec}$ is the probability of obtaining a solution for $m_t^{rec}$.

The dominant top mass dependence on the expected number of events comes from the theoretical $\sigma_{t\bar{t}}$. We use a NLO calculation of $\sigma_{t\bar{t}}$ evaluated at three different top masses \cite{xsec}; we parametrize the mass dependence of $\sigma_{t\bar{t}}$ on the top mass using the functional form described in \cite{fform}:

\begin{equation}
\sigma_{t\bar{t}}(M_{t})=6.70\cdot e^{(175-M_{t})/32.29}\ \rm{pb}.
\end{equation}

\noindent The acceptance $a$($M_{t}$) was studied using $t\bar{t}$ Monte Carlo simulation, separately for $b$-tagged and non-tagged samples. The Monte Carlo acceptances were corrected for trigger efficiencies and for scale factors arising from differences between data and simulation. The combined Monte Carlo acceptance corrections are between 74$\%$ and 95$\%$, depending on the lepton flavor and pseudorapidity. The dependence of the acceptance on the top mass is linear, increasing about 30$\%$ in the top mass range of 150 to 200~GeV/$c^2$. The integrated luminosity, ${L}$, is 1118~pb$^{-1}$ for the $b$-tagged sample and 1189~pb$^{-1}$ for the non-tagged sample. The two integrated luminosities are different because of the smaller data set using the silicon detector, which is required for $b$-tagging. The signal mass reconstruction probability, $p_{mass}^{rec}$, was measured to be 91$\pm1.1\%$ for both $b$-tagged and non-tagged samples, and was found to be independent of the top mass.

The cross section constrained top mass measurement uses information from the reconstructed top mass distribution as well as the observed number of events. The per-event mass reconstruction method and the template parametrizations are exactly the same as in the traditional measurement. The information from the number of events is added in the likelihood function by replacing $n_{\rm s}$ in Eq.~(\ref{eq_likelihood}) with  $n_{\rm s}(M_t)$ from Eq.~(\ref{eq_ns}); thus ${\cal L}  \equiv {\cal L}_{\rm b-tagged}(M_{t},n_{\rm b}^{\rm b}) \times {\cal L}_{\rm non-tagged}(M_{t},n_{\rm b}^{\rm non})$. The number of background events $n_{\rm b}$ and the top mass $M_{t}$ are free fit parameters as in the likelihood function of the traditional measurement. The uncertainty in the theoretical modeling of $\sigma_{t\bar{t}}$ is not included in the likelihood; it is treated in the same way as other systematic uncertainties described below.

Simulated experiments are used to verify that the cross section constrained method is unbiased and returns appropriate uncertainties. We measure $M_{t}=170.7^{+4.2}_{-3.9}$(stat.)~GeV/$c^2$. The statistical uncertainty is consistent with expectations: 34$\%$ of Monte Carlo experiments returned an uncertainty smaller than the uncertainty measured from data.

The sources of systematic uncertainties are summarized in Table~\ref{tab:sys}. The jet energy scale uncertainty is dominated by the uncertainty in jet energy corrections. This uncertainty was studied by shifting the jet energies by $\pm1\sigma$, and half of the mass difference was taken as the systematic uncertainty. Since the jet energy corrections were determined for light quark jets, we evaluated an additional systematic uncertainty from possible differences between $b$ jets and light quark jets \cite{bjes}. The total uncertainty from the jet energy scale is 1.8~(2.9)~GeV/$c^2$ for the cross section constrained (traditional) measurement.  The cross section constrained measurement is less sensitive to the jet energy corrections because a change in the jet energy scale shifts the top mass determination from the event yield in the opposite direction to that from kinematic reconstruction. The signal modeling uncertainty is 0.9~(0.8)~GeV/$c^2$, and takes into account differences in parton showering between the {\sc pythia}~\cite{pythia} and {\sc herwig}~\cite{herwig} Monte Carlo generators, uncertainties in initial and final state radiation modeling, and differences in parton distribution functions between MRST~\cite{mrst} and the full set of CTEQ6M~\cite{cteq} eigenvectors. Possible imperfections in modeling the $Z\rightarrow{ll}$ and misidentified lepton backgrounds combine to give 0.3~(0.3)~GeV/$c^2$ background modeling uncertainty. The uncertainty from template statistics is 0.4~(0.5)~GeV/$c^2$. A 1$\%$ uncertainty in the lepton $p_T$ introduces an uncertainty of 0.2~(0.2)~GeV/$c^2$. The cross section constrained measurement has an additional uncertainty of 1.6~GeV/$c^2$ from the expected number of events. This uncertainty includes 1.1~GeV/$c^2$ uncertainty from the integrated luminosity, 0.5~GeV/$c^2$ from the acceptance, 0.9~GeV/$c^2$ from the expected number of background events and 0.5~GeV/$c^2$ from the mass reconstruction probability.

\begin{table}[h]
\caption{Summary of systematic uncertainties for the traditional (T) and cross section constrained (C) measurements.}
\begin{center}
\begin{tabular}{l c } \hline
\hline
Systematic Source & $\Delta M_{t}$ (GeV/$c^2$) \\
\hline
& T\ \ \ \ \ C\\
Jet energy scale & 2.9\ \ \ \ 1.8 \\
Signal modeling & 0.8\ \ \ \ 0.9 \\
Background modeling & 0.3\ \ \ \ 0.3 \\
Template statistics & 0.5\ \ \ \ 0.4\\
Lepton $p_T$& 0.2\ \ \ \ 0.2 \\
Expected number of events & n.a\ \ \ \ 1.6\\
\hline
Total & 3.1\ \ \ \ 2.6\\
\hline
\hline
\end{tabular}
%\end{large}
\end{center}
\label{tab:sys}\rm
\end{table}

\begin{figure}[htbp]
\begin{center}
\includegraphics[width=0.4\textwidth]{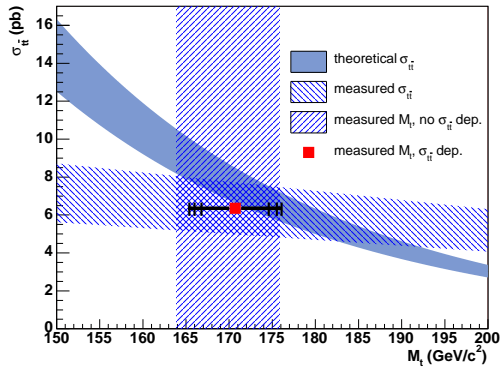}
\caption{The measured cross section constrained top mass is shown in the $M_{t}$~--~$\sigma_{t\bar{t}}$ plane. The innermost error bars correspond to the statistical uncertainty, the middle ones the statistical+systematic uncertainty, and the outermost error bars show the statistical+systematic+theory uncertainty. The hatched areas mark the traditional top mass measurement and the $\sigma_{t\bar{t}}$ measurement in the dilepton channel with statistical+systematic uncertainties.}
\label{fig:results}
\end{center}
\end{figure}

The uncertainty in the theoretical $\sigma_{t\bar{t}}(M_{t}=175~{\rm GeV}/c^2)$ is $^{+0.71}_{-0.88}$~pb \cite{xsec}. We propagated this uncertainty to the top mass by changing the number of signal events in the Monte Carlo experiments. The estimated uncertainty on the top mass is 2.4~$\rm{GeV/{\it{c}}^2}$. Simulation studies show that this cross section constrained top mass measurement is not very sensitive to the probability shape of the theoretical $\sigma_{t\bar{t}}$ uncertainty. Figure \ref{fig:results} shows the cross section constrained top mass measurement in the $M_{t}$~--~$\sigma_{t\bar{t}}$ plane. The extracted top mass from the cross section measurement~\cite{DIL_1.2} only is $\rm{178.3^{+10.1}_{-8.0}(exp)^{+4.0}_{-5.8}(theory)\ GeV/{\it{c}}^{2}}$, consistent within about one standard deviation with the result from the traditional analysis.

In summary, we have introduced a new way to improve the template top mass measurement in the dilepton channel by using a theoretical cross section constraint. With this measurement, we compare the reconstructed top mass distribution to templates and the observed number of events to expectation. In 1.2~fb$^{-1}$ of data collected by the CDF II detector, we measure a top quark mass of $\rm{170.7^{+4.2}_{-3.9}(stat)\pm2.6(syst)\pm2.4(theory)\ GeV/{\it{c}}^{2}}$. This measurement is in good agreement with the top mass measurement made without a cross section constraint, which gives $\rm{169.7^{+5.2}_{-4.9}(stat)\pm3.1(syst)\ GeV/{\it{c}}^{2}}$, and with top quark mass measurements made in other decay channels \cite{ljets,allhadronic,dzero_topmass}.

We thank the Fermilab staff and the technical staffs of the participating institutions for their vital contributions. This work was supported by the U.S. Department of Energy and National Science Foundation; the Italian Istituto Nazionale di Fisica Nucleare; the Ministry of Education, Culture, Sports, Science and Technology of Japan; the Natural Sciences and Engineering Research Council of Canada; the National Science Council of the Republic of China; the Swiss National Science Foundation; the A.P. Sloan Foundation; the Bundesministerium f\"ur Bildung und Forschung, Germany; the Korean Science and Engineering Foundation and the Korean Research Foundation; the Science and Technology Facilities Council and the Royal Society, UK; the Institut National de Physique Nucleaire et Physique des Particules/CNRS; the Russian Foundation for Basic Research; the Comisi\'on Interministerial de Ciencia y Tecnolog\'{\i}a, Spain; the European Community's Human Potential Programme; the Slovak R$\&$D Agency; and the Academy of Finland.


\begin{thebibliography}{99}
  \bibitem{higgs} 
  %The LEP Collaborations, LEP Electroweak Working Group, SLD Electroweak and Heavy Flavor Groups, CERN-PH-EP/2004-069.
  The LEP Collaborations, LEP Electroweak Working Group, SLD Electroweak and Heavy Flavor Groups, CERN-PH-EP/2006-042.
  %\bibitem{newphysics}
  %G. L. Kane and S. Mrenna, Phys. Rev. Lett. {\bf 77}, 3502 (1996).
  \bibitem{PRD}
  A. Abulencia {\it et al.} (CDF Collaboration), Phys. Rev. D {\bf 73}, 112006 (2006).
  \bibitem{madcow}
   A. Abulencia {\it et al.} (CDF Collaboration), Phys. Rev. D {\bf 75}, 031105 (2007).
  \bibitem{dzero_dil_mtop}
  V. M. Abazov  {\it et al.} (D0 Collaboration), Phys. Lett. B {\bf 655}, 7 (2007).
  \bibitem{xsec}
  M. Cacciari {\it et al.},  J. High Energy Phys. {\bf 04}, 68 (2004).
  \bibitem{fform}
  S. Catani {\it et al.}, Phys. Lett. B {\bf 378}, 329 (1996).
  \bibitem{CDF}
  D. Acosta {\it et al.} (CDF Collaboration), Phys. Rev. D {\bf 71}, 032001 (2005).
  \bibitem{ele_pt}
  For electrons, transverse energy is used.
  \bibitem{DIL}
  D. Acosta {\it et al.} (CDF Collaboration), Phys. Rev. Lett. {\bf 93}, 142001 (2004).
  \bibitem{btagging}
  T. Affolder {\it et al.} (CDF Collaboration), Phys. Rev. D {\bf 64}, 032002 (2001).
  \bibitem{pythia}
  T. Sjostrand {\it et al.}, Comput. Phys. Commun. {\bf 135}, 238 (2001).
  \bibitem{detsim}
  T. Affolder {\it et al.}, Nucl. Instrum. Methods Phys. Res. A{\bf 447}, 1 (2000).
  \bibitem{bjes}
   A. Abulencia {\it et al.} (CDF Collaboration), Phys. Rev. D {\bf 73}, 032003 (2006).
  \bibitem{herwig}
  G. Corcella {\it et al.},  J. High Energy Phys. {\bf 01}, 10 (2001).
  \bibitem{mrst}
  A. D. Martin {\it et al.}, Phys. Lett. B {\bf 356}, 89 (1995).
  \bibitem{cteq}
  J. Pumplin {\it et al.}, J. High Energy Phys.  {\bf 0207}, 012 (2002).
  \bibitem{DIL_1.2}
  M. Datta for the CDF and D0 Collaborations, proceedings of Hadron Collider Physics Symposium 2007.
  \bibitem{ljets}
  T. Aaltonen {\it et al.} (CDF Collaboration), Phys. Rev. Lett. (to be published), hep-ex/0703045.
  \bibitem{allhadronic}
  T. Aaltonen {\it et al.} (CDF Collaboration), hep-ex/0706.3790.
  \bibitem{dzero_topmass}
  V. M. Abazov  {\it et al.} (D0 Collaboration), Phys. Rev. D {\bf 74}, 092005 (2006).
\end{thebibliography}
\end{document}